# Precision measurement and determinations of laser radiation coherence by digital processing of their interferograms

E.A.Tikhonov, Seniour SPIE member, Institute of Physics
National Academy of Sci., Kiev, Ukraine, <etikh@live.ru>

*Abstract*—**The method of digital processing of continuous laser radiation interferograms by the digital microscope and original two-beam interferometer with the phase Bragg grating as the amplitude beam splitter for a quantitative determination of the spatial and temporary coherence module of radiation is described. The digital processing of the interferogram images was carried out with known software algorithms of OriginPro85. Principles and functionality of a measurement procedures are demonstrated on measuring of the radiation coherence of several gas, solid-state and diode lasers.**

I. INTRODUCTION REMARKS

In our publications [1,2] performances of phase Bragg grating (PBG) and configurations of a slit-free spectroscopic instrumentations on their base were considered. Two-beam interferometer on PBG, exercising the smooth and exact amplitude division of the input beam on two power equal was presented in work [3]. In paper [4] the simple justification of the digital processing and characterization of interferogram images were described.

 In the present contribution the procedure and results of spatial and temporary coherence measuring in the mentioned approach are provided. Here it is apropriate to remind that light coherence measurings remains an important part of the analysis of lasing mode content of traditional lasers,
 radiation of superluminous diodes, an undulatory X-ray rediation and correspondent methods are constantly improved /5-7/.

2. EXPERIMENTS AND MEASURING PROCEDURES

In proposed interferometer the two laser beams of equal intensity with the direct and reversed wavefronts form an interference pattern (IP) in a short-range diffraction area with a spatial period some mikrons in order do not exceed the resolution ability of used microscope with the digital eyepiece (TREK DCM320). Appeared IP is combined with the microscope observation plane. Typically the field of microscope observation much less than real IP size connected with the laser crossed beams. The visibility of observable part IP is measured quantitatively, averaged in whole or its smaller part, which is considered in future as the standard unit of coherence readout. To study the coherence distribution across whole IP the microscope scans interference pattern along a direction of the greatest visibility gradient. The perpendicular direction of IP do not use for scanning because along this direction visibility of IP does not change as connected with. zero phase difference of running beams . For the case of spatial coherence measuring the shoulder length of the interferometer should be taken equal, while at time coherence measuring in one of interferometer shoulders the adjustable delay line is set, but the interference angle stay nonvariable and nonzero.

Thus the proposed 2-beam interferometer allows to measure the time coherence observing in fact a variation of spatial interference pattern under condition of the variable difference of interferometer shoulder lengths: each point of a time delay here corresponds the delay time dependent distribution of the spatial coherence.

2D → 3D transform of IP and its numerical analysis (determination of the contrast relation and a local spatial period) were performed with the application of software OridginPro85 allowing also to carry out suppression of various noises in registered interferogram image. On Fig. 1. procedure of digital processing IP of the size 400x510 pixel is presented.

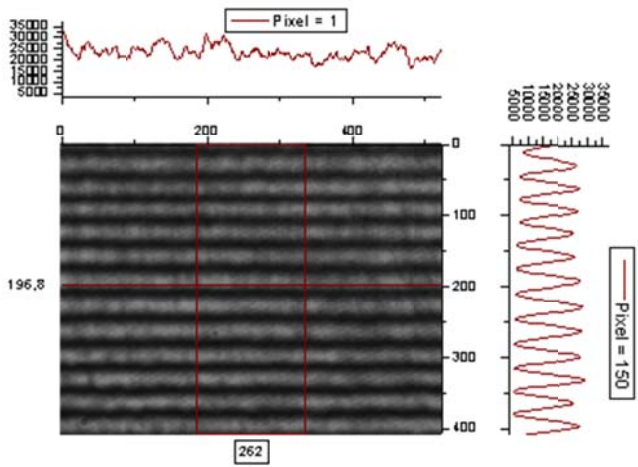

Fig.1. Luminosity distribution in a 16-bit code for longitudinal (top) and transversal (on the right) cross sections in width 1 and 150 pixel correspondently. Along x-y axes there are phase planes in pixels, along z-axis - luminosity of interference patterns in bits.

The picture Fig.1. shows process of pattern visibility measuring as in one point, so in the square of arbitrary sizes. If visibility values are homogeneous across the total observable image to choose the standard unit of measurement is not necessity. If visibility values across IP are essentially nonhomogeneous it is necessary to choose the smaller standard square with approximately equal visibility value with following averaging on all IP squares. At microscope scanning across the total IP to the next observable and treated IP square one must be adhere to the same coordinates of new observable image.

Averaging on an area 150x400pixel is shown on lateral (right) view Fig.1.

Next step of visibility processing and calculation consist in application of nonlinear regression method to estimate a degree of correlation between observable profiles of IP and simplified model imaginationheory. Theoretical dependence IP in the case of monochromatic (laser) light looks like **Y=const+sin$^2$(y)** so the degree of its correlation with experiment can be evaluated by the specified method.

When value of correlation occurs to be satisfied (parameter $R^2 \approx 1$), then calculated function parameters Y allow to find the magnitude of visibility and the spatial period of studied IP. On Fig.2A and 2B calculated and experimental results are presented for comparison.

Fig.2A.Graphic presentation of experimental

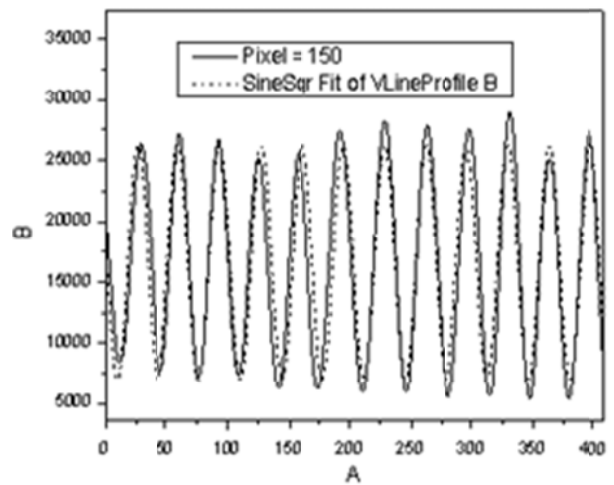

profiles of IP image and Y-function as approximation resulted from nonlinear regressin analysis.

Calculated parameters of Y-function allow to determine magnitude of visibility and coherence of IP on the following formula that results from the standard definition of contrast including arbitrary noises:

| Noise level | $Y_0$ | 6961,4 | 231,75 |
|---|---|---|---|
| Shift of max | $x_c$ | -24,56 | 0,227 |
| Spatial period in pixel | w | 33,72 | 0,029 |
| Amplitude, bit | A | 19278,7 | 379,48 |
| $Y=Y_0+A \sin^2(\pi(x-x_c)/w)$ | | | an error |

Fig.2B- table with calculated parameters and errors for Y-function as approximation of experimental datas.

$$K_n=A/(A+Y_0) \qquad (1)$$

In considered case the spatial period of IP w=33,72 pixel, contrast magnitude results from above mentioned formula $K_n=A/(A+Y_0)$ = 19279/(19279+6961)=73,5 %. Taking into account the level of dark noises of used camera equal ≈2750bit we can improve accuracy of obtained results.

With application of the featured methods measuring coherence module of the several gas, diode and solid-state pumped by diode laser have been determined.

The expected case of spatial coherence distribution was registered for **one transverse mode** of He-Ne the laser. The coherence of an extended coherent light source of the size l with wave vector k apart of given source **r** can be described by expression [8]

$$V(x)=2|\sin(kxl/2r)/kxl/2r|/[1+(\sin(kxl/2r)/kxl/2r)^2] \qquad (2)$$

and for given He-Ne laser presented on Fig.3.:

From Fig.3 follows that size of spatial coherence on distance r=300mm has increased from 300mm has increased from 0,1mm up to ≈2,5mm (50% level)

Fig.4,5. show results of coherent measurement of the many-mode He-Ne in dependence of a pinhole diameter that was installed before analysed laser beam. This operation provided the partial removing of the transverse modes of the studied laser. Results of coherence measurement that are shown on Fig.4. were obtained without any pinhole on the laser beam. They brings the weak similarity with calculated coherence behavior presented on Fig.3. Similarity in the same distribution increased considerely when the output pinhole 1mm is installed (fig.5.).

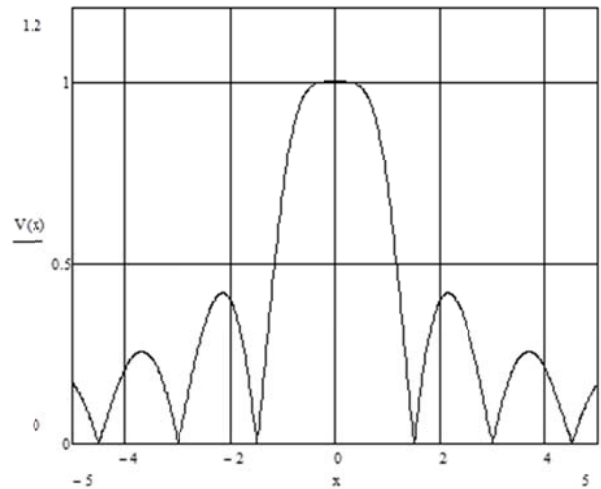

Fig.3. Distribution of spatial coherence for one-mode He-Ne laser for r =300mm, l= 0,1mm.

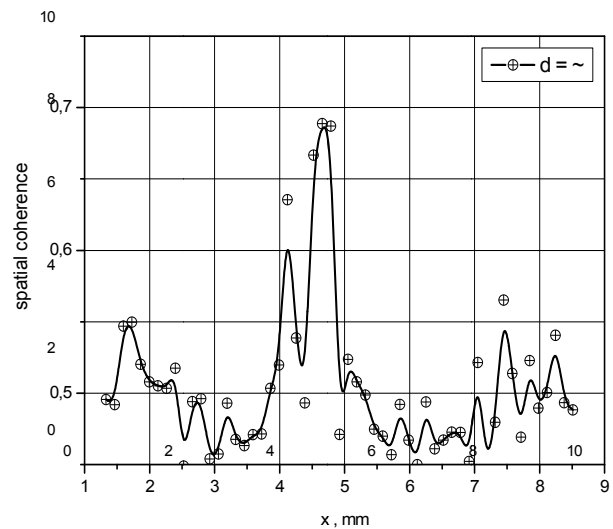

Fig.4. Spatial coherence distribution for the many-mode He-Ne laser without spatial filtration

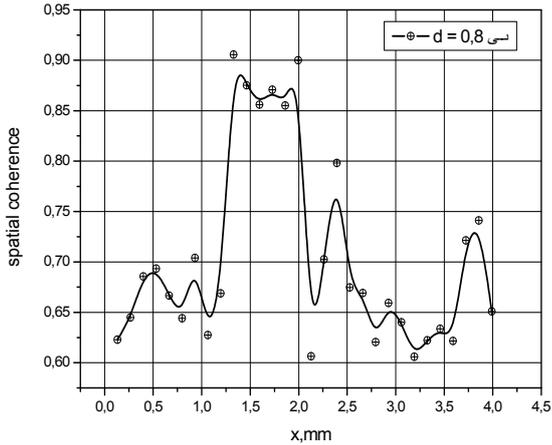

Fig.5. Spatial coherence distribution the same laser with spatial pinhole ∅1mm.

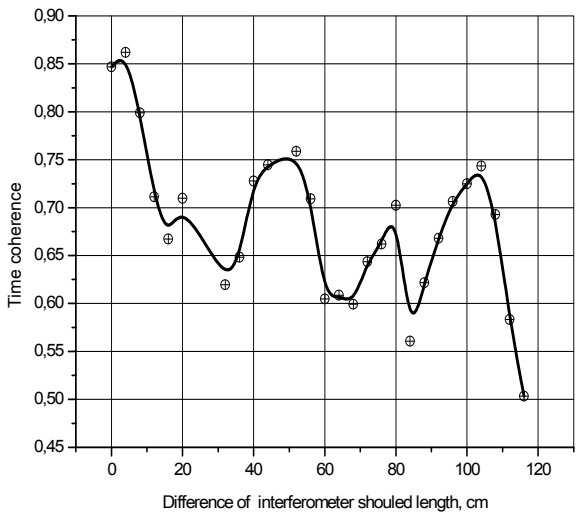

Fig.6. Time coherence measurement for two-mode He-Ne laser

Dependence for time coherence of 1-mode He-Ne laser was measured with the same interference angle as for above used configurations (see Fig.6.) that is why the obtained coherence distribution remembers dependence on the picture on fig.5. Moreover length of coherence for He-Ne laser is more than used delay line 1,2m. therefore results of the experiment do not discover strong effect of time coherence variation.

Similar coherence measurements discovered also unexpectedly high coherence magnitudes for the usual diode laser, but detail discussion of it spreads beyond limits this paper.

## FINAL REMARKS

Two-beam interferometer with volume holographic phase grating as the new diffraction Bragg optical element of the scheme with a smooth 2-beam splitting and tuning of interference beams relative to amplitudes and phase difference was carried out. The digital method of recording and processing interferogram images have been preposed to allow rapid and precise measurement of visibilaty contrast and spatial period for an interference field and, accordingly, the coherence of radiation most types of lasers with precision, which remained elusive with the previous photodensitometry techniques.

Digital registration and processing of interferogram images using existing digital image processing allows to begin the new stage of coherence laser measurements and research. The usage of digital microscopes enables to provide the similar image analysis of interferograms in a wide range of spatial frequencies.